\documentclass[aps,prc,twocolumn,superscriptaddress,altaffilletter,nobibnotes,nofootinbib,10pt,showpacs,showkeys,preprintnumbers]{revtex4-2}

\usepackage{float}
\usepackage{graphicx}
\usepackage{mathptmx}
\usepackage{flushend}
\usepackage{dcolumn}
\usepackage[dvipsnames]{xcolor}
\usepackage{bm}
\usepackage{lineno}
\usepackage{float}
\usepackage{amsmath}
\usepackage{soul}
\usepackage{multirow}
\usepackage{tikz,xcolor,hyperref}

\definecolor{lime}{HTML}{A6CE39}
\DeclareRobustCommand{\orcidicon}{
	\begin{tikzpicture}
	\draw[lime, fill=lime] (0,0) 
	circle [radius=0.16] 
	node[white] {{\fontfamily{qag}\selectfont \tiny ID}};
	\draw[white, fill=white] (-0.0625,0.095) 
	circle [radius=0.007];
	\end{tikzpicture}
	\hspace{-2mm}
}
\foreach \x in {A, ..., Z}{\expandafter\xdef\csname orcid\x\endcsname{\noexpand\href{https://orcid.org/\csname orcidauthor\x\endcsname}
			{\noexpand\orcidicon}}
}

\newcommand{\pte}{$p_{\rm{_T}}$}
\newcommand{\pt}{$p_{\rm{_T}}$~}

\newcommand{\pb}{$PbPb$~}

\begin{document}


\title{Model comparison of the transverse momentum spectra of charged hadrons produced in \pb collision at $\sqrt{s_{NN}} = 5.02$ TeV}
\author{Rohit Gupta}
\affiliation{Department of Physical Sciences, Indian Institute of Science Education and Research (IISER) Mohali, Sector 81 SAS Nagar, Manauli PO 140306 Punjab, India}

\author{Satyajit Jena}
\email{sjena@iisermohali.ac.in}
\affiliation{Department of Physical Sciences, Indian Institute of Science Education and Research (IISER) Mohali, Sector 81 SAS Nagar, Manauli PO 140306 Punjab, India}

\begin{abstract}
Transverse Momentum, \pte, spectra is of prime importance in order to extract crucial information about the evolution dynamics of the system of particles produced in the collider experiments. In this work, the transverse momentum spectra of charged hadrons produced in \pb collision at $5.02$ TeV has been analyzed using different distribution functions in order to gain strong insight into the information that can be extracted from the spectra. We have also discussed the applicability of unified statistical framework on the spectra of charged hadron at $5.02$ TeV.
\end{abstract}

\maketitle
\section{Introduction}
The quest to develop the understanding of Quantum Chromodynamics (QCD) matter created during the very early universe has been the primary motivation behind several theoretical as well as experimental studies in particle physics. Whether the QCD phase transition is first order, second order or simple crossover, and what is the critical temperature and the order parameters of phase transition etc.~are some of the long-standing questions that intrigue the researchers to explore QCD phase diagram by utilising data from different heavy-ion collider experiments like Relativistic Heavy Ion Collider (RHIC) at BNL and in Large Hadron Collider (LHC) at CERN.  The QCD matter under discussion is popularly known as Quark-Gluon Plasma (QGP) which cooks up under extreme conditions of temperature and energy density. When the temperature $(T)$ and the energy density $(\epsilon)$ reaches a critical value, quark no longer remain confined inside their individual hadron and instead become free to move over a nuclear volume forming a deconfined QGP state. Such extreme condition has been achieved by colliding heavy ions moving at the ultra-relativistic speed in the experiments at RHIC \& LHC and provided us the tool to investigate and to uncover the mystery of the fireball that filled our universe few microseconds after the Big Bang. 

The theoretical reasoning behind the formation of QGP comes from the asymptotic freedom due to the nature of the running coupling strength of constituent partons. The theory of quark gluon interaction, QCD, dictates that the coupling strength between quarks \& gluons  increases with a decrease in the energy scales (or increase in length scale).  On the other hand, the length scale can be decreased by bringing partons closer to each other. As the length scale reduces, the coupling strength also decreases and it reaches to a point where quarks \& gluons asymptotically appear to be free from their nucleonic volume changing its hadronic degrees of freedom to partonic degrees of freedom. Hence a dense hot soup of quarks and gluons forms, which we call QGP.  This is exactly what experiments like RHIC and LHC had achieved by colliding highly energetic nucleons travelling at ultra-relativistic speed.  However, the time scale of such process is extremely small so it is not possible to directly probe and characterize it using present day technologies. In experiments, we only receive the information about kinematics of final state particles that are free streaming to the detectors. We utilise the kinematics of these final state particles to get the information about the underlying process that led to the formation of these particles. 

The transverse momentum, \pte, spectra of final state particle is a crucial kinematic observable that can be utilized as an effective probe to understand the thermodynamical properties of the system produced in the collision. 
It is also an essential observable to understand the dynamics of QGP and the quark-hadron phase transition.
 Several theoretical models \cite{Stodolsky:1995ds, Tsallis:1987eu, Ristea:2013ara, Tang:2008ud} have been developed to characterize \pte-spectra and to extract different physical parameters that can be further used to enhance our understanding of the system produced during such collisions. Most of these models are developed based on the statistical and thermodynamical approach since it is almost impossible to apply perturbative field theory calculations because of the high QCD coupling strength at low \pt (due to asymptotic freedom). 
 
In this work, we have performed a comparative study of different models on \pte-spectra of charged hadrons produced in \pb~collision at the recently published result at $\sqrt{s_{NN}}~=~5.02$ TeV \cite{Acharya:2018qsh} along with the $2.76$ TeV \cite{Abelev:2012hxa} data measured by the ALICE experiment at CERN. 
\section{Models}
As discussed in the previous section, it is difficult to apply perturbative QCD  at low-\pt region of spectra as the coupling strength becomes extreemly high at low momentum transfer scale. Hence, we rely on the phenomenological models. Among several such models,  statistical thermal approaches are widely used to explain transverse momentum spectra. Initial work by Koppe \cite{Koppe:1948, Tawfik:2013tza}, Fermi \cite{fermi:ptp, Fermi:1951zz} and Hagedorn \cite{Hagedorn:1965st, Hagedorn:1967ua} developed the fundamentals of the statistical description of particle spectra in high energy physics. These initial works was followed by several application of statistical thermodynamics in studying the particle production mechanism using the momenta carried out by outgoing particles.
In this section, we will discuss different models developed to explain the spectra and we will present the results obtained by fitting the \pte-spectra data produced in $5.02$ TeV collision. Some of the models discussed below requires the information of mass of the constituent particle. Since a larger contribution of charged hadrons comes from pion ($\sim 70\%$), we have used the mass of pion while analyzing the spectra of charged hadrons.

\subsection{Boltzmann Distribution}
For a system that is considered to be of purely thermal origin, the most natural choice to explain the statistical distribution of the energy of constituent particles is the standard Boltzmann-Gibbs statistics.
Boltzmann distribution has long been used to explain energy spectra \cite{Schnedermann:1993ws, Stodolsky:1995ds} of the classical statistical system. Because the temperature of the system during heavy-ion collision is extremely high and Fermi-Dirac and Bose-Einstein statistics approaches Maxwell-Boltzmann statistics at high temperature, we can use Boltzmann distribution to explain the spectra of particles produced during the heavy-ion collision. In the case of Boltzmann distribution, we can write number density as \cite{Schnedermann:1993ws, Stodolsky:1995ds}:
\begin{equation}\label{BG}
\frac{d^2N}{2\pi p_T dp_T dy}=m_T \frac{gV}{(2\pi)^3}\; exp\left(\frac{-m_T}{T} \right)
\end{equation}  
\begin{figure}[!h]
\centering
   \includegraphics[height=2in,width=3in]{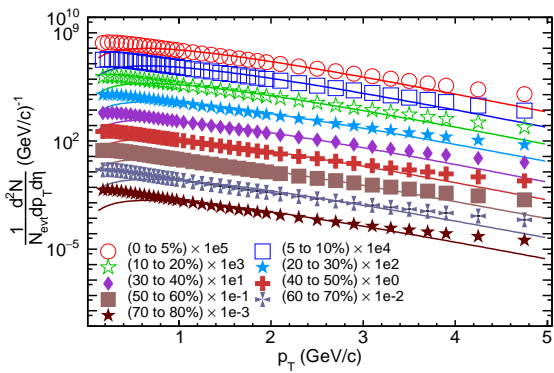}
  \caption{(color online) The transverse momentum spectra of charged hadrons at different centralities produced at collision energies of 5.02 TeV \cite{Acharya:2018qsh} fitted with Boltzmann (Eq.~\ref{BG}).}
  \label{fig:Boltzmann Fit276}
\end{figure} 
Here $m_T$ is the transverse mass and is related to transverse momentum by $m_T = \sqrt{m^2 + p_T^2}$, $g$ is spin degeneracy factor and $V$ is volume of the system under consideration. 
In Fig.~\ref{fig:Boltzmann Fit276} we have fitted the \pte-spectra of charged hadrons produced \pb collision at $5.02$ TeV with the Boltzmann distribution function (Eq.~\ref{BG}). We observe that the data deviate significantly from the BG function at low- \& high-\pt region of the spectra. This deviation finds its origin in the low number of particles in the collision. BG statistics applies to the system where the number of particles in the system should be of the order of Avogadro number; however, only a few thousand particles are produced in heavy-ion collision. At the same time, the applicability of BG statistics is limited to the system where entropy of the ensemble is extensive as well as additive in nature. To tackle these fundamental issues, we need to find a solution for non-extensive formalism which can be applied to the system under discussion. Therefore, a generalization of BG statistics was proposed by C. Tsallis in his seminal work Ref.~\cite{Tsallis:1987eu} to tackle non-extensivity in the system. By doing so it overcomes the limitation of BG statistics. In this context, we provide a brief discussion on Tsallis statistics and the corresponding fit result of \pb collision at $\sqrt{s_{NN}}~=~5.02$ TeV. 
\subsection{Tsallis Distribution}

To provide a generalized formalism to explain non-extensive system such as the system produced in the heavy-ion collision, Tsallis statistics, also known as non-extensive statistics, were proposed as a generalization of Boltzmann-Gibbs statistics in 1988 by C. Tsallis. In Tsallis statistics, a parameter $'q'$ has been introduced, which takes care of the non-extensivity in the system. This new parameter also acts as a scaling factor of number of particle to make this statistics applicable to the system with the low number of particles compared to the Avogadro number. Here, normal exponential are replaced by q-exponential as
\begin{equation}
exp_q(x) = \left[ 1 - (q-1)x\right] ^{-\frac{1}{q-1}}
\end{equation}  
which in the limit $q\rightarrow 1$ gives us normal exponential.

Further, entropy in case of Tsallis statistics \cite{Tsallis:1987eu} is given as
\begin{equation}
S_q = k \frac{1-\sum_i p_i^q}{q-1}
\end{equation}
\begin{figure}[!h]
\centering
  \includegraphics[height=2in,width=3in]{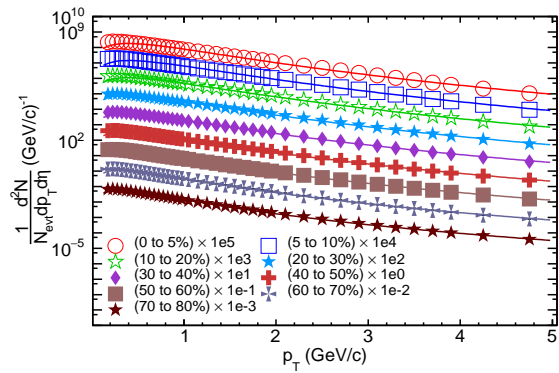}
  \caption{(color online) The transverse momentum spectra of charged hadrons at different centralities produced at collision energies of 5.02 TeV \cite{Acharya:2018qsh} fitted with Tsallis (Eq.~\ref{tsallis}).}
  \label{fig:Tsallis Fit}
\end{figure}
Functional form of transverse momentum spectra in case of Tsallis statistics can be obtained by replacing normal exponential in Eq.~(\ref{BG}) with q-exponential
\begin{equation}\label{tsallis}
\frac{1}{2\pi p_T} \frac{d^2 N}{dp_T dy} = \frac{gV m_T}{(2 \pi)^3} \left[1+(q-1)\frac{m_T}{T}\right]^{-\frac{q}{q-1}} 
\end{equation}
This distribution function converges to Boltzmann distribution in the limit $q\rightarrow 1$. Also, Tsallis statistics has been proved to be thermodynamically consistent following the laws of thermodynamics \cite{Cleymans:2012ya}.
This form of transverse momentum spectra has been shown in different work \cite{Wilk:2001db, Cleymans:2012ya, Wilk:2012zn, Cleymans:2011in, Khuntia:2017ite} to nicely explain the spectra in limited \pt region. Additionally, it has been observed that there are certain systems \cite{Reference27} with significant long-range interactions, the entropy in such system can be non-additive and non-extensive and the Tsallis formalism has been used to explain such system \cite{Tsallis:2009zex}.

Figure \ref{fig:Tsallis Fit} shows the fitting of Tsallis distribution to the  \pte-spectra of charged particles produced in \pb collision at $\sqrt{s_{NN}}~=~5.02$ TeV. It is evident from the figure that Tsallis distribution provides a much better fit to the spectra compared to the Boltzmann distribution. The detail comparison of $\chi^2/NDF$ values are given in the table \ref{table_1}, we can observed that there is a significant improvement in explaining the spectra of \pb collision at $\sqrt{s_{NN}}~=~5.02$ TeV. 

However, we observed that it doesn't explain the data at low- and high-\pt part of spectra. This might be due to the various particle production mechanism in the heavy-ion collision, effecting the nature of spectra. Therefore, the applicability of Tsallis statistics is only limited to soft-\pt regime of the spectra (although it deviates narrowly at low-\pte) whereas for high-\pte, dominated by hard-processes, it doesn't explains. As shown and discussed in Ref.~\cite{Wong:2013sca, Saraswat:2017kpg, Azmi:2015xqa}, the Tsallis formalism becomes extremely complex to incorporate hard processes and eventually a modification is required to  explain very high-\pt part of the particle spectra. On the contrary, the particle production is dominated by hard QCD processes at high-\pt region of the spectra, which could be explained by a QCD inspired power law form of function such as:
\begin{equation}
\label{eqn:pqcd}
f(p_T) = \frac{1}{N} \frac{dN}{dp_T} = Ap_T\left(1 + \frac{p_T}{p_0} \right)^{-n} 
\end{equation}
An uninfied model to explain the contribution of both soft and hard processes to \pte-spectra is still an open problem. Before we discuss about the unified formalism \cite{Jena:2020wno, Gupta:2020naz}, we would like to discuss about several other approaches to explain \pte-spectra.  Apart from the purely statistical models discussed above, several hydrodynamics based models are also developed to explain the spectra considering the initial state fluctuation and the collective behaviour of the QCD matter created during the collision.
\subsection{Blast Wave (BW)}
It is experimentally established the system produced during the high energy collision has an azimuthal anisotropy because of the difference in flow velocities along different direction. This azimuthal anisotropy is a result of some initial state geometrical effects that arise during the collision. Thus, it can be argued that the outgoing particles must carry the imprint of such effects and can influence the nature of the spectra. To incorporate such effects in spectra, several models are proposed \cite{Ristea:2013ara, Tang:2008ud}. The Blast wave model is a hydrodynamics inspired model developed to include the flow properties along with the random thermal motion of particles in order to give a complete picture of the evolution dynamics of QGP, including the azimuthal anisotropy. Transverse momentum spectra in case of Blast-Wave model \cite{Ristea:2013ara} is given as :
\begin{equation}\label{BWEqn}
\begin{aligned}
\frac{dN}{p_Tdp_T}\propto \int_0^R r\; dr\; m_T\; & I_0 \left( \frac{p_T\; sinh\; \rho(r)}{T_{kin}}\right) \\ 
& \times K_1\left( \frac{m_T\; cosh\; \rho(r)}{T_{kin}}\right) 
\end{aligned}
\end{equation} 
Here $m_T$ is the transverse mass, $I_0$ and $K$ are modified Bessel functions. Also, $\rho (r) = tanh^{-1}\beta$ and $\beta$ is the transverse radial flow velocity which is parameterized in form of a power law of the form $\beta_t(r) = \beta_s(r/R)^n$ where $\beta_s$ represent the surface flow velocity and $n$ is the exponent of flow profile. The average transverse flow velocity $(\langle \beta_T \rangle )$ is given in term of $\beta_s$ and exponent $n$ as $\langle \beta_T \rangle = \beta_s\times \frac{2}{2+n}$ \cite{Ristea:2013ara}. Finally, $r/R$ represents the radial position of thermal source. \\
\begin{figure}[!h]
\centering
  \includegraphics[height=2in,width=3in]{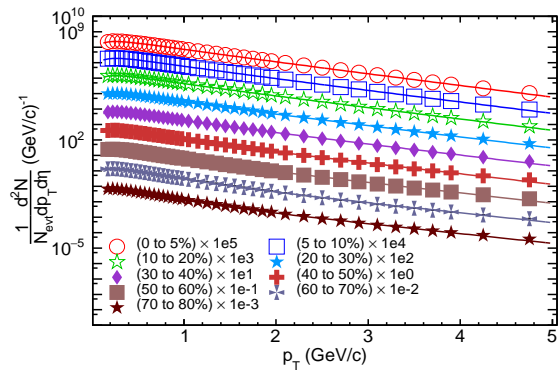}
  \caption{(color online) The transverse momentum spectra of charged hadrons at different centralities produced at collision energies of 5.02 TeV \cite{Acharya:2018qsh} fitted with Blast-Wave function (Eq.~\ref{BWEqn}).}
  \label{fig:BW Fit}
\end{figure}
The temperature extracted using the BW function has a different interpretation as compared to the temperature extracted from Boltzmann-Gibbs function (Eq.~\ref{BG}) or Tsallis function (Eq.~\ref{tsallis}). Fitting with BW function gives us kinetic freeze-out temperature $(T_{kin})$ whereas fitting with BG or Tsallis distribution gives us effective temperature $(T_{eff})$. These two quantities are related as \cite{Basu:2015zra}:
\begin{equation}
T_{eff} = T_{kin}+f(\beta_t)
\end{equation} 
Here $f(\beta_t)$ is a function of transverse flow velocity $\beta_t$.

We show the result of Blast-Wave fitting to the \pte-spectra of particles produced in $5.02$ TeV collision in the Fig.~\ref{fig:BW Fit}. The fitting yields the kinetic temperature as $113.6$ MeV, and the average flow velocity close to $0.63c$ for most central collision. The $\chi^2/NDF$ values are close to unity for central collisions but they start to increase with the decrease in the nuclear overlap by approaching to a value 20 for most peripheral collision. This formalism doesn't include the concept of non-extensive, therefore, a generalization of Blast-Wave model to include non-extensive system is discussed in the next section.
\subsection{Tsallis Blast Wave (TBW)}
Tsallis Blast Wave method is an extension of Blast-Wave method which was introduced to take into account the effect of non-extensivity in the system along with the flow properties. As discussed elsewhere in the article, Tsallis formalism can be used to tackle the system with non-extensivity, the Boltzmann distribution used in the Blast-Wave model is replaced by the Tsallis distribution in order to get the Tsallis Blast Wave function \cite{Ristea:2013ara, Tang:2008ud}. Transverse momentum spectra in the case of TBW is given as:
\begin{equation}\label{TBWEqn}
\begin{aligned}
& \frac{dN}{p_Tdp_T} \propto m_T \int_{-Y}^{Y} cosh(y)\; dy \int_{-\pi}^{\pi} d\phi \int_0^R rdr \times \\ 
& \left( 1 + \frac{q-1}{T} (m_T\; cosh(y)\; cosh(\rho)-p_T\;sinh(\rho)\; cos\phi)\right)^{\frac{-1}{q-1}}
\end{aligned}
\end{equation}
Here $y$ represent rapidity and $\rho$ is the flow profile as described in the Blast-Wave section. 

After its introduction, the TBW method has been used in different works to fit \pte-spectra of particles and extract information related to the thermodynamics and hydrodynamics evolution of the system produced in the heavy-ion collision. For TBW fit, the value of $n$ is set to one \cite{Ristea:2013ara, Tang:2008ud}.
 \begin{figure}[!h]
\centering
  \includegraphics[height=2in,width=3in]{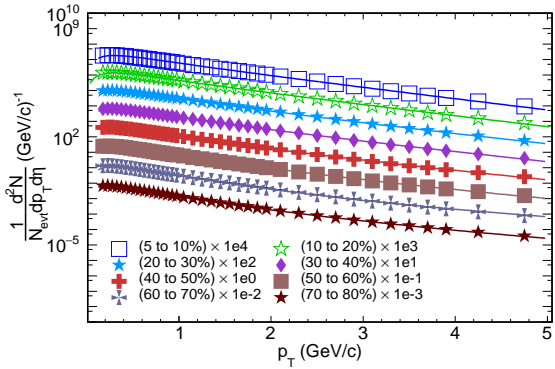}
  \caption{(color online) The transverse momentum spectra of charged hadrons at different centralities produced at collision energies of 5.02 TeV \cite{Acharya:2018qsh} fitted with Tsallis Blast-Wave function (Eq.~\ref{TBWEqn}).}
  \label{fig:TBW Fit}
\end{figure}
\begin{table*}[t]
\centering
\caption{Table of parameter values obtained after fitting charged hadron spectra at 3 different centralities with different functions}
\label{table_1}
\begin{tabular*}{\textwidth}{@{\extracolsep{\fill}}lrrrrl@{}}
\hline
\multicolumn{1}{@{}l}{Variable} & Method & Energy (TeV)&$0$ to $5\%$&$40$ to $50\%$&$70$ to $80\%$\\ 
\hline
 \multirow{6}{*}{$T_{eff}(MeV)$} &\multirow{2}{*}{BG}&$2.76$&$316.95\pm2.74$&$316.3\pm3.58$&$305.23\pm5.75$\\ 
   &&$5.02$&$319.747\pm0.67$&$330.97\pm0.67$&$338.522\pm0.68$ \\ \cline{2-6}
  &\multirow{2}{*}{Tsallis}&$2.76$&$163\pm3.995$&$137.52\pm3.839$&$111.22\pm4.113$\\ 
   &&$5.02$&$176.101\pm1.049$&$148.433\pm0.775$&$117.75\pm0.640$ \\ \cline{2-6}  
    &\multirow{2}{*}{Unified function}&$2.76$&$393.535\pm44.567$&$311.339\pm77.432$&$296.085\pm215.735$\\ 
   &&$5.02$&$407.448\pm3.322$&$369.097\pm18.529$&$329.4\pm34.236$ \\ \hline
 \multirow{4}{*}{$T_{kin}(MeV)$} &\multirow{2}{*}{BW}&$2.76$&$124.25\pm67.877$&$163.18\pm1.700$&$157.73\pm11.591$\\ 
   &&$5.02$&$113.571\pm14.841$&$157.637\pm4.268$&$164.249\pm1.847$ \\ \cline{2-6}
 &\multirow{2}{*}{TBW}&$2.76$&$76.983\pm5.939$&$52.006\pm6.026$&$21.143\pm20.156$\\ 
   &&$5.02$&$-$&$76.292\pm1.062$&$44.363\pm1.163$ \\ \hline
 \multirow{8}{*}{q} &\multirow{2}{*}{Tsallis}&$2.76$&$1.0945\pm0.0027$&$1.1186\pm0.0029$&$1.138\pm0.0034$\\ 
   &&$5.02$&$1.09736\pm0.00072$&$1.12023\pm0.00057$&$1.1416\pm0.00052$ \\ \cline{2-6}
 &\multirow{2}{*}{TBW}&$2.76$&$1.0135\pm0.0102$&$1.0548\pm0.0088$&$1.0967\pm0.0178$\\ 
   &&$5.02$&$-$&$1.04678\pm0.00231$&$1.08609\pm0.00147$ \\ \cline{2-6}   
  &\multirow{2}{*}{$q-weibull$}&$2.76$&$1.0212\pm0.0141$&$1.0613\pm0.0166$&$1.0905\pm0.0218$\\ 
   &&$5.02$&$1.00274\pm0.00409$&$1.04755\pm0.00346$&$1.08525\pm0.00359$ \\ \cline{2-6} 
   &\multirow{2}{*}{Unified function}&$2.76$&$1.0478\pm0.0038$&$1.0935\pm0.0084$&$1.1325\pm0.0312$\\ 
   &&$5.02$&$1.04808\pm0.00018$&$1.08476\pm0.00177$&$1.1315\pm0.00317$ \\ \hline
 \multirow{12}{*}{$\chi^2/NDF$} &\multirow{2}{*}{$BG$}&$2.76$&$25.345$&$29.819$&$24.0907$\\ 
   &&$5.02$&$396.783$&$735.272$&$1080.25$ \\ \cline{2-6} 
 &\multirow{2}{*}{$Tsallis$}&$2.76$&$1.994$&$1.1226$&$0.51061$\\ 
   &&$5.02$&$34.3003$&$36.3687$&$23.6161$ \\ \cline{2-6} 
 &\multirow{2}{*}{$BW$}&$2.76$&$0.296$&$0.603$&$0.525$\\ 
   &&$5.02$&$2.08721$&$12.0452$&$20.6522$ \\ \cline{2-6}         
  &\multirow{2}{*}{$TBW$}&$2.76$&$0.666$&$0.515$&$0.351$\\ 
   &&$5.02$&$-$&$4.74803$&$1.74155$ \\ \cline{2-6}   
 &\multirow{2}{*}{$q-weibull$}&$2.76$&$0.11$&$0.059$&$0.018$\\ 
   &&$5.02$&$1.41606$&$2.19311$&$1.15846$ \\  \cline{2-6}
&\multirow{2}{*}{Unified function}&$2.76$&$0.101$&$0.052$&$0.017$\\ 
   &&$5.02$&$1.70914$&$1.96482$&$1.08821$ \\ 
\hline
\end{tabular*}
\end{table*}
We have shown the fitting of \pte-spectra with TBW function in Fig.~\ref{fig:TBW Fit}. The value of kinetic freeze-out temperature extracted from the TBW fit shows a decline trend with the increase in centrality and the average flow velocity is around $0.4c$. TBW model gives a holistic view of particle production in high energy collision, including statistical and hydrodynamical description but only at the low-\pt part of the spectra. Although the result obtained by this method gives a better explanation at low-\pte, it does not take care of hardness in the spectra in high-\pt region, which is dominated by the particles produced in hard scattering processes. 

We also tried to fit this data with statistical model as proposed in Ref.~\cite{Dash:2018qln}.
The q-Weibull distribution is an extension of Weibull distribution, which was described by Swedish mathematician Waloddi Weibull in 1951. Weibull distribution is a continuous probability distribution and is given as:
\[
P(x,\lambda,k)=
\begin{cases}
\frac{k}{\lambda} \left( \frac{x}{\lambda}\right)^{k-1}\; e^{-(x/\lambda)^k} & x \geq 0\\
0 & x<0
\end{cases}
\]
where $\lambda$ is the scale parameter and $k$ is the shape parameter of distribution, with $k$ \& $\lambda > 0$. 

Weibull distribution finds its application in the system where dynamical evolution is driven by fragmentation and sequential branching \cite{brown1989, brown1995derivation}. Since the evolution of the system in hadrons and heavy-ion collision is dominated by a perturbative QCD based parton cascade model, we can apply $q$-Weibull distribution to study particle spectra. 
\begin{figure}[!h]
\centering
  \includegraphics[height=2in,width=3in]{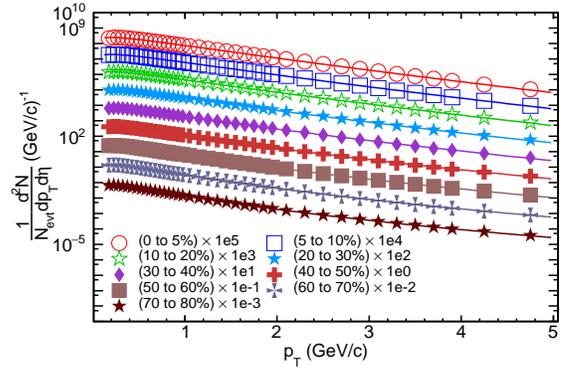}
  \caption{(color online) The transverse momentum spectra of charged hadrons at different centralities produced at collision energies of 5.02 TeV \cite{Acharya:2018qsh} fitted with q-Weibull function (Eq.~\ref{weibullform}).}
  \label{fig:qweibull Fit}
\end{figure}
Incorporating Tsallis framework in Weibull distribution gives us the q-Weibull distribution \cite{Dash:2018qln},
\begin{equation}\label{weibullform}
P_q(x,q,\lambda,k)=\frac{k}{\lambda} \left( \frac{x}{\lambda}\right)^{k-1}\; e_q^{-(\frac{x}{\lambda})^k}
\end{equation}
where 
\begin{equation}\label{weibulexp}
e_q^{-(\frac{x}{\lambda})^k} = \left( 1-(1-q)\left( \frac{x}{\lambda}\right)^k\right)^{\left( \frac{1}{1-q} \right)}   
\end{equation}
In Fig.~\ref{fig:qweibull Fit} we have performed $q$-Weibull fit to the transverse momentum spectra data and the fit show good agreement with the data. 
Its relevance in particle spectra lies in its ability to fit spectra over a wide range of \pt as compared to the other models described above with good accuracy. 
The models discussed so far are mostly described the data at low-\pt quite well as they do not include the hard QCD processes in their construction. In the next section we will discuss a unified formalism that describe both soft- and hard-processes in a consistent manner.

\subsection{Unified Distribution function}
Pearson distribution is a generalized probability distribution, which, under different limit on its parameter, reduces to different distribution function like exponential, Gaussian, Gamma distribution, Student's t-distribution etc. It was introduced in 1895 by Karl Pearson in his seminal work \cite{Pearson343}. It has been applied quite successfully in different fields such as geophysics, statistics and financial marketing. It has been introduced in the context of particle production in heavy-ion collision for the first time in Ref.~\cite{Jena:2020wno}.
\begin{figure}[!h]
\centering
  \includegraphics[height=2in,width=3in]{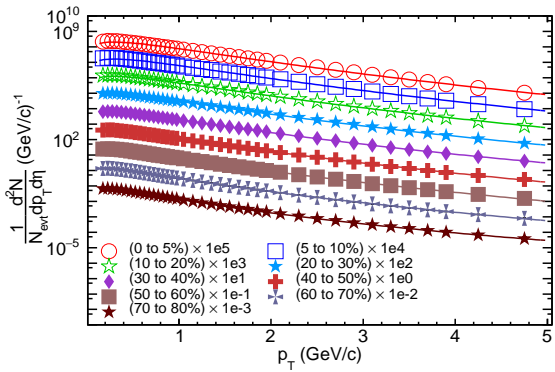}
  \caption{(color online) The transverse momentum spectra of charged hadrons at different centralities produced at collision energies of 5.02 TeV \cite{Acharya:2018qsh} fitted with unified distribution function.} 
  \label{fig:pearson Fit}
\end{figure}
The transverse momentum spectra in case of unified distribution is given as 
\begin{equation}
 \label{eq:pearson_final}
\frac{1}{2\pi p_T} \frac{d^2 N}{dp_T dy} =  B' \left( 1 + \frac{p_T}{p_0}\right)^{-n} \left( 1 + (q-1)\frac{p_T}{T}\right)^{-\frac{q}{q-1}}
\end{equation}
As it has been described in Ref.~\cite{Jena:2020wno, Gupta:2020naz}, this form of \pt distribution function nicely takes care of both the ``hard" and ``soft" part of the spectra. Further, the unified formalism also provide a connection to the flow of particle as discussed in Ref.~\cite{Jena:2020wno}. 

Figure \ref{fig:pearson Fit} represents a good agreement between transverse momentum spectra  and the unified distribution function in the form of Eq.~(\ref{eq:pearson_final}). Apart from providing the best fit to the transverse momentum spectra (as is evident from the $\chi^2/NDF$ values provided in table  \ref{table_1}), unified distribution is also more physical in the sense that it is thermodynamically consistent \cite{ Gupta:2020naz} and it is also proved to be backward compatible to Tsallis distribution under limiting conditions. 

\section{Result}
We have performed the fitting of transverse momentum spectra using different \pt distribution function like Boltzmann-Gibbs function, Tsallis distribution function, Blast-Wave, Tsallis Blast-Wave, q-Weibull and unified framework. We have used the recently released \pte-spectra data of charged hadrons produced in \pb~collision at $5.02$ TeV measured by ALICE experiment \cite{Acharya:2018qsh}. For comparison, we have also performed the similar analysis on the charged hadrons \pte-spectra data produced in $2.76$ TeV \pb~collision \cite{Abelev:2012hxa}. 

In the above section, we have provided the fitting plot for $5.02$ TeV \pte-spectra data. Fitting has been carried for different collision centrality classes including $0-5\%$ (most central collision), $40-50\%$ (mid-central collision) and $70-80\%$ (peripheral collision). We have used the ROOT \cite{Brun:1997pa} data analysis framework to perform the fittings.\\

\begin{figure}
\centering
  \includegraphics[height=2in,width=3in]{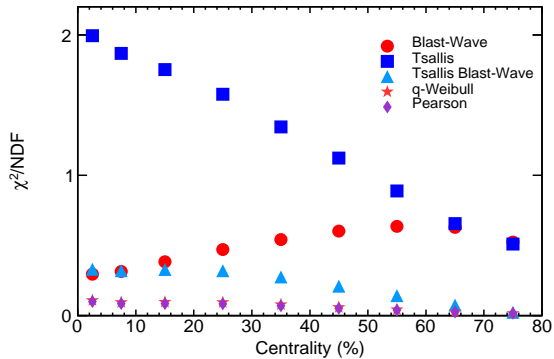}
  \caption{(color online) Value of $\chi^2/NDF$ obtained for different functions fitted with \pte-spectra data of particles produced at $2.76$ \pb~collision.}
  \label{fig:chi2_val_2760}
\end{figure}

\begin{figure}
\centering
  \includegraphics[height=2in,width=3in]{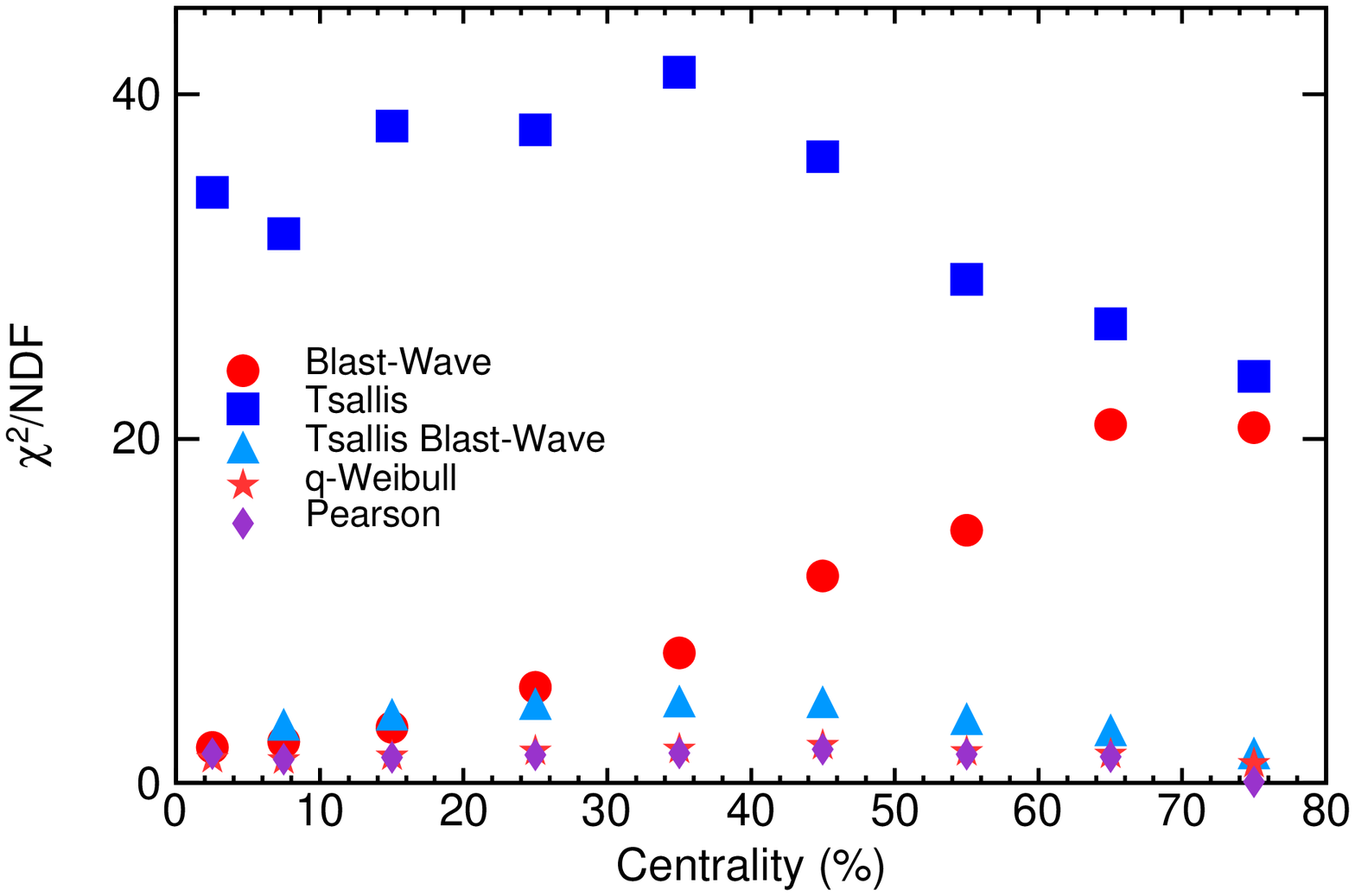}
  \caption{(color online) Value of $\chi^2/NDF$ obtained for different functions fitted with \pte-spectra data of particles produced at $5.02$ \pb~collision.}
  \label{fig:chi2_val_5020}
\end{figure}

\begin{figure}
\centering
  \includegraphics[height=2in,width=3in]{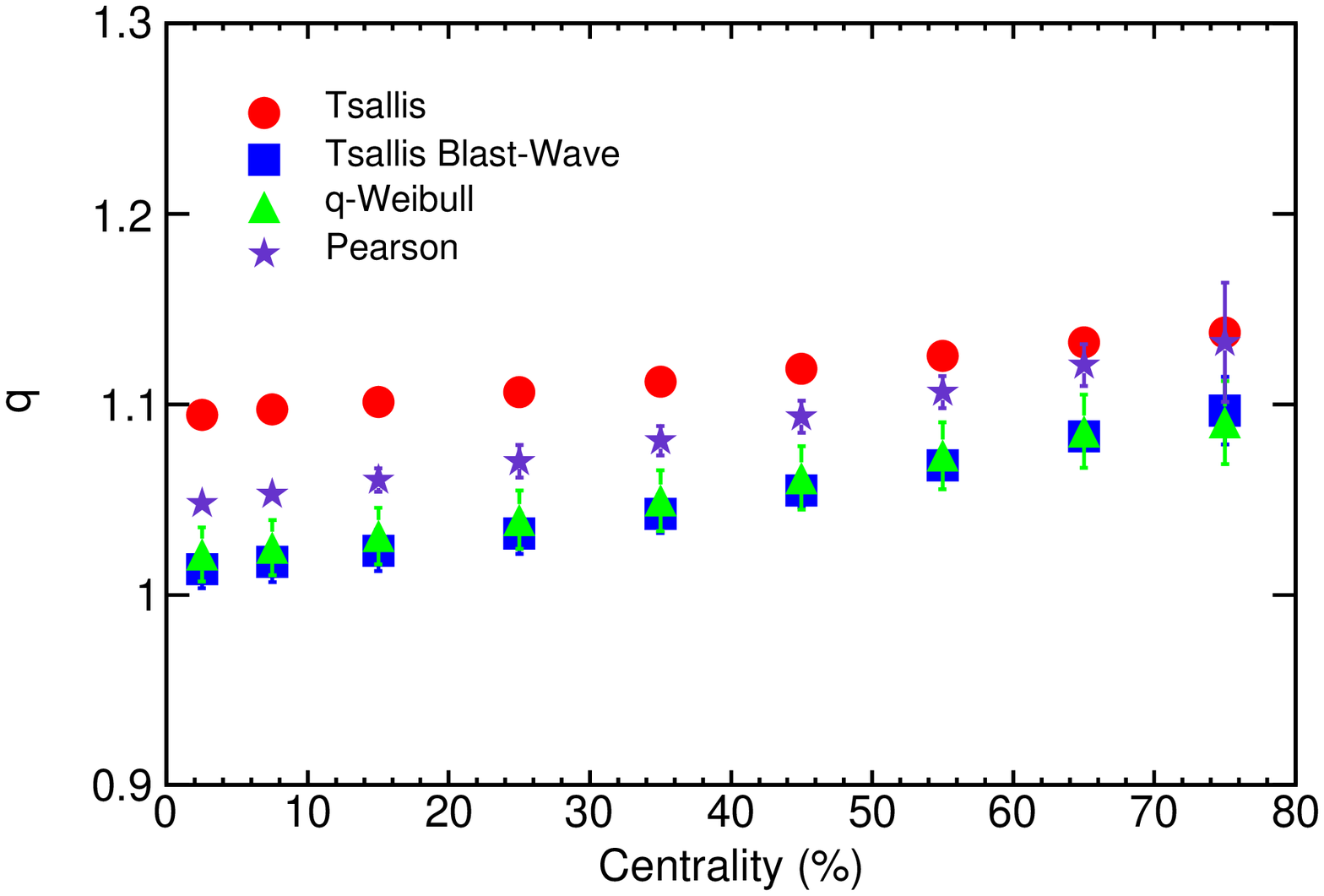}
  \caption{(color online) Value of non-extensivity parameter `q' for different functions fitted with \pte-spectra data of particles produced at $2.76$ TeV \pb~collision.}
  \label{fig:q_val_2760}
\end{figure}
\begin{figure}
\centering
  \includegraphics[height=2in,width=3in]{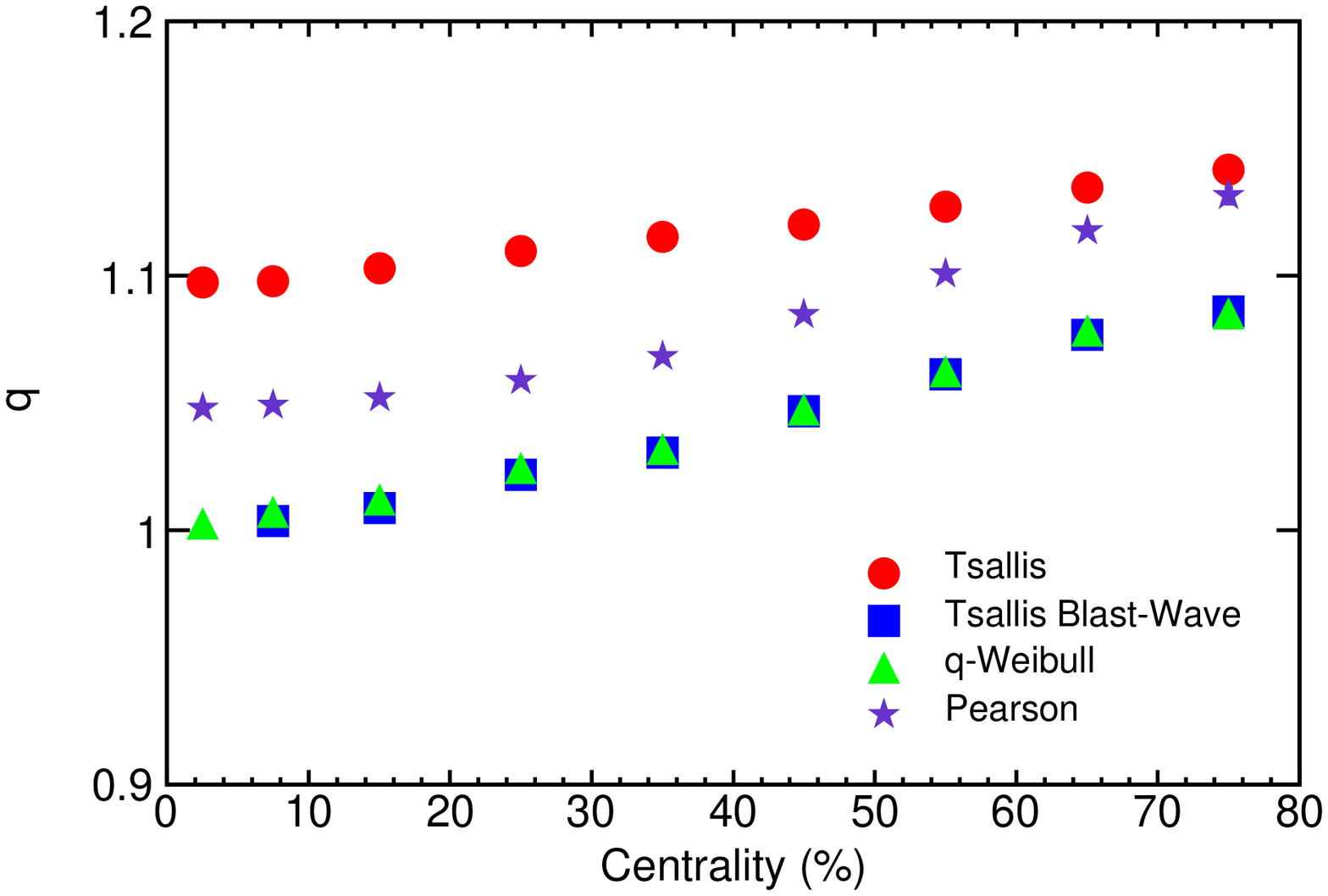}
  \caption{(color online) Value of non-extensivity parameter `q' for different functions fitted with \pte-spectra data of particles produced at $5.02$ TeV \pb~collision.}
  \label{fig:q_val_5020}
\end{figure}

In table \ref{table_1} and \ref{tab:BWTBW} along with figures \ref{fig:chi2_val_2760}, \ref{fig:chi2_val_5020}, \ref{fig:q_val_2760} and \ref{fig:q_val_5020}, best fit values of different parameters are presented along with the $\chi^2/NDF$ value which is a measure of the goodness of fit.
 
\begin{table*}[t]
\centering
 \caption{\label{tab:BWTBW}The best fit value of exponent $`n$' and average transverse flow velocity obtained by fitting the charged hadron transverse momentum spectra using BW and TBW models. }
\begin{tabular*}{\textwidth}{@{\extracolsep{\fill}}l|ll|ll|ll@{}}
\hline
\multirow{2}{*}{Centrality} & \multicolumn{2}{c|}{$n$ (BW)} &
\multicolumn{2}{c|}{$\langle \beta_T \rangle $ (BW)} &
\multicolumn{2}{c}{$\langle \beta_T \rangle $ (TBW)} 
\\ 
\cline{2-7}
 & $2.76$ TeV & $5.02$ TeV & $2.76$ TeV & $5.02$ TeV & $2.76$ TeV & $5.02$ TeV \\
\hline
\multirow{2}{*}{0 to 5 \%} & $1.2633 $ & $0.9471 $&$0.5544$&$0.6303$&$0.4241$&-  \\
&$\pm 0.7737$ & $\pm 0.1236$ & $\pm0.1442$ & $\pm0.0291$ & $\pm0.0048$ &\\
\hline
\multirow{2}{*}{5 to 10 \%} & $1.3660$ &$0.9538$ &$0.534$&$0.6316$&$0.4244$&$0.4181$ \\
& $\pm0.6137$ & $\pm0.1322$ & $\pm0.1059$ & $\pm0.0308$ & $\pm0.0048$ & $\pm0.0014$\\
\hline
\multirow{2}{*}{10 to 20 \%} & $1.5316$ & $0.9740$&$0.5047$&$0.6302$&$0.4247$&$0.4192$ \\
&$\pm1.6441$ & $\pm0.1465$ & $\pm0.2528$ & $\pm0.0334$ & $\pm0.0052$ & $\pm0.0013 $\\
\hline
\multirow{2}{*}{20 to 30 \%} & $1.8038$&$1.1591$&$0.4642$&$0.5862$&$0.4254$&$0.4186$\\
&$\pm0.5546$ & $\pm0.1704$ & $\pm0.0714$ & $\pm0.0338$ & $\pm0.0052$ & $\pm0.0014$\\
\hline
\multirow{2}{*}{30 to 40 \%} & $2.1756$&$1.4738$&$0.4199$&$0.5225$&$0.4263$&$0.4199$\\
&$\pm0.4673$ & $\pm0.2004$ & $\pm0.0487$ & $\pm0.0319$ & $\pm0.0054$ & $\pm0.0012$\\
\hline
\multirow{2}{*}{40 to 50} \% & $2.6499$&$1.9766$&$0.3771$&$0.4479$&$0.4281$ &$0.4189$\\
&$\pm0.4699$ & $\pm0.0814$ & $\pm0.0389$ & $\pm0.0095$ & $\pm0.0048$ & $\pm0.0014$\\
\hline
\multirow{2}{*}{50 to 60 \%} & $3.1212$ & $2.4192$&$0.3446$&$0.4033$&$0.4305$&$0.4200$\\
& $\pm0.4993$ & $\pm0.0799$ & $\pm0.0341$ & $\pm0.0075$ & $\pm0.0051$ & $\pm0.0016$\\
\hline

\multirow{2}{*}{60 to 70 \%} & $3.6574$ & $3.2153$&$0.3149$&$0.3424$&$0.4330$&$0.4209$ \\
& $\pm0.5502$ & $\pm0.0877$ & $\pm0.0309$ & $\pm0.0054$ & $\pm0.0052$ & $\pm0.0016$ \\
\hline
\multirow{2}{*}{70 to 80 \%} &  $3.9250$ & $3.4730$&$0.3041$&$0.3301$&$0.4355$&$0.4257$ \\
& $\pm0.5907$& $\pm0.0797$ & $\pm0.0305$ & $\pm0.0048$ & $\pm0.0083$ & $\pm0.0013$\\
\hline
\end{tabular*}
  
\end{table*}
  
   \begin{table*}
\centering
\caption{\label{tab:qWeibull}The best fit value of parameters $k$ and $\lambda$ obtained by fitting the charged hadron transverse momentum spectra using $q$-Weibull model. }
\begin{tabular*}{\textwidth}{@{\extracolsep{\fill}}lll|ll@{}}
\hline
\multirow{2}{*}{Centrality} & \multicolumn{2}{c|}{$k$} &
\multicolumn{2}{c}{$\lambda$} 
 
\\ 
\cline{2-5}
 & $2.76$ TeV & $5.02$ TeV & $2.76$ TeV & $5.02$ TeV\\
\hline
0 to 5 \% & $0.8183 \pm 0.0481$ & $0.7666 \pm 0.0127$&$0.1953\pm0.0170$&$0.1974\pm0.0049$  \\
5 to 10 \% & $0.8218\pm0.0493$ &$0.7741\pm0.0132$ &$0.1949\pm0.0172$&$0.1988\pm0.0049$ \\
10 to 20 \% & $0.8297\pm0.0503$ & $0.7788\pm0.0118$&$0.1940\pm0.0172$&$0.1972\pm0.0044$ \\
20 to 30 \% & $0.8407\pm0.0520$&$0.7998\pm0.0115$&$0.1910\pm0.0171$&$0.1968\pm0.0040$\\
30 to 40 \% & $0.8545\pm0.0546$&$0.8035\pm0.0110$&$0.1865\pm0.0172$&$0.1894\pm0.0037$\\
40 to 50 \% & $0.8684\pm0.0574$&$0.8295\pm0.0111$&$0.1791\pm0.0171$&$0.1853\pm0.0035$\\
50 to 60 \% & $0.8816\pm0.0621$ & $0.8492\pm0.0119$&$0.1703\pm0.0175$&$0.1772\pm0.0035$\\
60 to 70 \% & $0.8971\pm0.0690$ & $0.8717\pm0.0156$&$0.1611\pm0.0182$&$0.1678\pm0.0032$ \\
70 to 80 \% &  $0.8909\pm0.0783$ & $0.8662\pm0.0123$&$0.1489\pm0.0197$&$0.1536\pm0.0033$ \\
\hline
\end{tabular*}
  
\end{table*}
 Looking at the $q$-values, we can predict the non-extensivity in the system or, in other words, we can quantify how much a system deviates from the equilibrium. Large q values indicate a non-equilibrium system and the system approaches equilibrium as the q value approaches to unity.
 
  From table \ref{table_1}, $q$ values from all four methods, Tsallis, TBW, q-Weibull and unified function shows a decreasing trend as we move from most peripheral to most central collision indicating that the system is highly non-equilibrium in the peripheral collision and it moves toward equilibrium as the overlap region between two colliding nuclei increases. The trend in this result shows a similar pattern reported  previously for different energies in Ref.~\cite{Tang:2008ud,Ristea:2013ara,Dash:2018qln}.
  
  Table \ref{tab:BWTBW} shows a decreasing trend in the value of average transverse flow velocity estimated by BW model as we go from central to peripheral collision. The corresponding numerical values for TBW fit does not show any centrality dependence and also the values are lower as compared to the BW model. This difference can be attributed to the absence of the non-equilibrium description in the BW model as it demands the system to be in thermal equilibrium.
  
   Apart from the non-extensivity parameter, the $q$-Weibull fit also provide the value of two more free parameters $\lambda$ $\&$ $k$. The best fit value of these parameter is presented in the table \ref{tab:qWeibull}. The value of parameter $k$ increases as we go from central to peripheral collision for both the energies. The values are also consistently higher in $2.76$ TeV as compared to the $5.02$ TeV. The parameter $\lambda$ show a reverse trend with the values decreasing from central to peripheral collision. The physics interpretation of these parameters are still an open problem and a detail study of $q$-Weibull function is required to have a better understanding of the  model.
   
 Among all different distribution functions, we have obtained the best fit for the unified distribution, which is also evident from the $\chi^2/NDF$ values. One possible reason for the low $\chi^2/NDF$ in unified distribution as compared to other distribution is that it also takes care of hard perturbative QCD processes, which affect the \pt distribution at higher \pt range as described in Ref.~\cite{Jena:2020wno}. Another interesting result which are under-discussed, is related to the scaling properties of $q$ parameter. Since, the $q$ parameter acts as a scaling factor to apply statistical mechanics at the low number of particles,  the number of final state particles decreases with the increase in centrality and hence the corresponding scaling parameter should increase with centrality. We observe this in figures \ref{fig:q_val_2760}, \ref{fig:q_val_5020} for all four different methods indicating that the $q$ parameter can be interpreted as a scaling factor depending upon the number of final state particles in the system.
 \section{Conclusion}
 In this work, we have analyzed the \pte-spectra of charged particles produced in \pb collision at $\sqrt{s_{NN}}~=~5.02$ TeV using six different formalism which are based on different physics inputs for example, Boltzmann distribution is a purely thermal model, Blast-Wave models also include collective flow effect and the unified distribution which also takes into account hard QCD processes. The effects such as non-extensivity and collective flow that need to be taken into account in a model for explaining such data. From the variation of q-parameter with centrality, we also observe that system in highly non-equilibrium at the peripheral collision and moves toward the equilibrium as we move toward the most central collision. Further, a very high value of $\chi^2/NDF$ indicates that a purely thermal model is not a good explanation for the \pt spectra.  Apart from this, the scaling properties of the $q$ parameter have also been established, indicating its relevance in small systems.
 
  In conclusion, we can say that a complete model to explain transverse momentum spectra must include the non-extensivity and collective flow on top of random thermal motion and should also be able to explain the effects on high-\pt range arising due to hard pQCD processes. The essence of all these physics input is present in the unified distribution and it also provide the best fit to the transverse momentum spectra as is evident from the values of $\chi^2/NDF$ presented in Fig.~\ref{fig:chi2_val_2760} \& \ref{fig:chi2_val_5020}.

\section{Data Availability}
The data used for this analysis are already published and are cited at relevant places within the text as references.

\section{Conflicts of Interest}
The authors declare that they have no known competing financial interests or personal relationships that could have appeared to influence the work reported in this paper.

\section*{Acknowledgement}
  R. Gupta would like to acknowledge the financial support provided by CSIR through fellowship number 09/947 (0067) 2015-EMR-1.

\end{document}